\documentclass[prl,twocolumn,amsmath,amssymb,superscriptaddress,showpacs,numbers,openany]{revtex4-2}

\usepackage[utf8]{inputenc}
\usepackage{color}
\usepackage{latexsym}
\usepackage{float,ulem}
\usepackage{dsfont}
\usepackage[dvipsnames]{xcolor}
\usepackage[english]{babel}
\usepackage{lipsum}

\usepackage{bm, mathtools}
\usepackage{graphicx}% Include figure files
\usepackage{dcolumn}% Align table columns on decimal point
\usepackage{times}
\usepackage[english]{babel}
\usepackage[T1]{fontenc}

\usepackage{cellspace}%
\usepackage{makecell}
\setcellgapes{3pt}

\newcommand{\dotnice}[1]{\overset{\,\bm.}{#1}\,{\vphantom{#1}}}

\definecolor{linkcolor}{rgb}{0,0,0.6} %hyperlink
\usepackage[pdftex,colorlinks=true, pdfstartview=FitV, linkcolor= linkcolor, citecolor= linkcolor, urlcolor= linkcolor, hyperindex=true,hyperfigures=true]{hyperref} %hyperlink% or \documentclass[page-classic]{epl2} for one column style

\begin{document}

\title{Unraveling active baths through their hidden degrees of freedom}

\author{D.M. Busiello}
\thanks{These authors equally contributed to this work}
\author{M. Ciarchi}
\thanks{These authors equally contributed to this work}
\author{I. Di Terlizzi}
\thanks{These authors equally contributed to this work}
\affiliation{Max Planck Institute for the Physics of Complex Systems, N{\"o}thnitzer Stra{\ss}e 38, 01187, Dresden, Germany}

%\date{\today}

\begin{abstract}
\noindent 

The dynamics of a probe particle is highly influenced by the nature of the bath in which it is immersed. In particular, baths composed by active (e.g., self-propelled) particles induce intriguing out-of-equilibrium effects on tracer's motion that are customarily described by integrating out the dynamics of the bath's degrees of freedom (DOFs). %However, thermodynamic quantities are usually severely affected by coarse-graining procedures.
However, thermodynamic quantities, such as the entropy production rate, are generally severely affected by coarse-graining procedures. Here, we show that active baths are associated with the presence of \textit{dissipative} DOFs exhibiting non-reciprocal interactions with a probe particle. Surprisingly, integrating out these DOFs inevitably results into a system-dependent increase or reduction of the entropy production rate. On the contrary, it stays invariant after integrating out \textit{non-dissipative} DOFs. As a consequence, they determine the dimensionality of isoentropic hypersurfaces in the parameter space. Our results shed light on the nature of active baths, revealing that the presence of a typical correlation time-scale is not a sufficient condition to have non-equilibrium effects on a probe particle, and draws a path towards the understanding of thermodynamically-consistent procedures to derive effective dynamics of observed DOFs.

%Active baths for example, which may be composed of self-propelled particles or agents, induce intriguing and peculiar out of equilibrium effects on the motion of the tracer. The equation of motion for the latter are usually obtained by coarse graining, or by "integrating out", the bath's degrees of freedom hence obtaining an effective description for the dynamics of the probe particle. While coarse graining preserves the system's dynamical features, the same does not hold for thermodynamic proprieties. In particular, entropy production, a key thermodynamic quantity measuring the irreversibility of an active system, is known to decrease after coarse graining. In this paper, we show how for Langevin systems, exact coarse graining of linear degrees of freedom (DOFs), denominated as \textit{entropic DOFs} can lead to both a decrease or an increase of entropy production. These DOFs can be then identified as the equivalent of an active bath. On the other side, the effects of a passive bath can be described by \textit{non etropic DOFs}, whose coarse graining leaves the global entropy production invariant. We will start from the latter case to illustrate the framework and to set the ground for the more involved part regarding active baths.

\end{abstract}

\maketitle

Biological systems are active, meaning that they are maintained out of equilibrium by internal or external energy-consuming processes \cite{fodor2022irreversibility,julicher2018hydrodynamic}. Bacteria \cite{elgeti2015physics}, colloidal particles with catalytic surfaces \cite{palacci2013living}, and enzymes \cite{ghosh2021enzymes} are a few prominent examples of active systems. The presence of this activity induces a variety of fascinating phenomena, such as anomalous transport \cite{granek2022anomalous,liang2022emergent}, swarming \cite{jin2021collective}, pattern formation \cite{bois2011pattern}, and motility-induced phase separation \cite{cates2015motility}.

In numerous experimental and theoretical settings, active systems act as environments for other observed particles \cite{maes2020fluctuating,pietzonka2017entropy}. This is the case of a passive tracer to probe the activity of an unknown medium \cite{banerjee2022tracer}, or a large fluorescent molecule in a solution of chemical buffers \cite{caspi2000enhanced,xu2019direct}. At variance with the classical description of equilibrium reservoirs, active baths do not necessarily relax faster to their steady-state distribution, introducing an additional time-scale which is usually entangled with the one characterizing the dynamics of the probe particle alone. This characteristic requires, in principle, the simultaneous description of all degrees of freedom (DOFs), making necessary a proper coarse-graining procedures to study how hidden (unobservable) variables affect observed ones. A popular framework aiming at capturing the effects induced by baths with a typical time-scale is the Generalized Langevin Equation (GLE) \cite{mori1965transport,kubo1966fluctuation}, an extension of the customary Langevin Equation to systems with memory kernels. A consistent thermodynamic approach compatible with this framework is a developing field \cite{mai2007nonequilibrium,OurGLE,di2020thermodynamic}, with the main issue being that GLEs provide only an effective description of the entire system.

Alternatively, a possible approach to obtain a thermodynamically complete description of active systems is to start from the set of equations describing all DOFs undergoing non-equilibrium processes, whether hidden or observables, using an effective stochastic description only for fast variables that form an equilibrium bath in which all the others are immersed. This approach is demanding but has a twofold advantage. First, all DOFs are Markovian; second, thermodynamic properties can be unambiguously obtained \cite{sekimoto1998langevin}. Surely enough, the dynamic of the probe particle is obtained by integrating out all hidden DOFs. However, the thermodynamics of non-equilibrium systems is severely affected by any coarse-graining procedure \cite{esposito2012stochastic,busiello2019entropy,yu2021inverse}, and no general schemes can circumvent this problem \cite{busiello2020coarse,busiello2019entropy2,ghosal2022inferring}.

Since both these approaches lead to dead-ends, it is difficult to establish the validity of either of the two. Analogously, it is generally impossible to quantify the dissipation of a probe particle in an active bath. Even more critically, it is also not known to date when the dynamics of a given bath of hidden slow variables modifies the dissipation of a probe particle, i.e., when the bath can be defined as active in a thermodynamically-consistent way.

Here, we tackle exactly this problem, showing that there exist a class of stochastic DOFs, that we name \textit{dissipative}, whose effect cannot be integrated out in a thermodynamically-consistent way, i.e., without affecting (decreasing or enhancing) the total entropy production rate of the system. These DOFs are associated to non-reciprocal interactions and are accountable for non-equilibrium effects. Conversely, \textit{non-dissipative} DOFs can be coarse-grained without modifying the dissipation. Our derivation unravels the nature of an active bath by inspecting its hidden DOFs, ascribing the origin of an active bath to the presence of non-reciprocal interactions between hidden and observed variables. Crucially, in the presence of an active bath, descriptions in terms of GLEs are thus never thermodynamically consistent, and one needs to retain all the DOFs involved in the dynamics.

As a starting point, we consider a set of two linear coupled stochastic differential equations for two DOFs, $x$ and $y$:
\begin{equation} \label{eq:2dsystem}
    \begin{split}
        \dot{x}_t &= f_t^x + A_{11}x_t+A_{12}y_t+\sqrt{2D}\xi^x_t \\
        \dot{y}_t &= f_t^y + A_{22}y_t+A_{21}x_t+\sqrt{2D}\xi^y_t
    \end{split}
\end{equation}
or written in matrix form: $\dot{\pmb{x}}_t = \pmb{F}_t + \sqrt{2\pmb{D}}\cdot\pmb{\xi}_t$, where the components of $\pmb{F}_t$, $F_t^x$ and $F_t^y$, are the total forces acting only along $x$ and $y$, respectively. Here, $f^x_t$ and $f^y_t$ are external non-conservative forces. In Eq.~\eqref{eq:2dsystem}, we already considered the friction coefficients to be equal to $1$ to keep the notation simple. Due to the fluctuation-dissipation theorem, the resulting factors in front of the noise terms have to be identical, i.e., $\sqrt{2D}$ in this case. In full generality, the presence of two different temperatures can always be mapped into an effective non-conservative force, e.g., for the Brownian gyrator \cite{filliger2007brownian,young2020nonequilibrium}. At any rate, all the results presented herein equally hold in the presence of different frictions and diffusion coefficients, as long as the fluctuation-dissipation theorem is fulfilled. The system of equation \eqref{eq:2dsystem} describes the mutual influence of two DOFs undergoing a stochastic motion while being coupled to
%due to the influence of an}
an external equilibrium bath.

The rate of entropy production in a non-equilibrium steady state (NESS) is related to the amount of heat dissipated in the environment and is given by \cite{sekimoto1998langevin}:
\begin{equation}
\label{sekimoto}
\sigma_{\rm tot} = \langle \pmb{F}_t\circ \dot{\pmb{x}}_t\rangle \;,
\end{equation}
taking a unitary diffusion coefficient without loss of generality. In the simple case of the system in Eq.~\eqref{eq:2dsystem}, $\sigma_{\rm tot} = \sigma_x + \sigma_y$, with $\sigma_a = \langle F^a_t \circ \dot{a}_t \rangle$, $a = x,y$. Notice that the $\sigma_x$ and $\sigma_y$ can be either positive or negative, as they quantify the independent contributions of the two DOFs to $\sigma_{\rm tot}$. When some DOFs are not observable, they effectively behave like a bath for the observed DOFs. To develop a better understanding of how non-equilibrium properties change in these scenarios, we study in detail the rate of entropy production.

Considering again Eq.~\eqref{eq:2dsystem}, we focus on the case in which $f_t^y = 0$ and $f_t^x$ is a generic non-conservative force. Moreover, we assume that both $A_{11}$ and $A_{22}$ are negative and the coupling terms are symmetric, $A_{12} = A_{21}$, so that the only source of non-equilibrium is $f_t^x$. In this scenario, we first notice that the total stationary entropy production rate coincides with the one of $x$ only \cite{supplemental_material}:
\begin{gather}\label{diss}
\sigma_{tot} = \langle \pmb{F}_t \circ \dot{\pmb{x}}_t \rangle = \langle f_t^x \circ \dot{x}_t \rangle = \sigma_x
%= \langle {\color{red}F_t^x} \circ \dot{x}_t \rangle ~+ \\
%+ ~\frac{A_{12}}{2} \dot{C}_{xy}(0) + \frac{A_{22}}{2} \dot{C}_{yy}(0) = \langle {\color{red}f_t^x} \circ \dot{x}_t \rangle = \sigma_x \nonumber \, ,
\end{gather}
Indeed, since the couplings between $y$ and $x$ are reciprocal, we define $y$ as a \textit{non-dissipative} DOF as it does not contribute to the entropy production rate of the whole system.

%{\color{blue}riscrivere formula meglio} where $C_{\alpha \beta}(t-s) = \langle \alpha_t \beta_s \rangle$ are the correlation functions. 

Consider now $x_t$ as the observed DOF. We can explicitly solve the equation for $y_t$ and replace this time-dependent solution into the evolution of $x_t$, hence obtaining an exponentially correlated colored noise \cite{supplemental_material}.
%Consider now $x$ as the observed DOF. We can explicitly solve the equation for $\dot{y}_t$, and replace the time-dependent solution $y(t)$ into the evolution of $x$. This substitution leads to a colored noise with an exponential memory kernel \cite{supplemental_material}.
Since we know that the fluctuation-dissipation theorem holds in our system, we reconstruct the dissipation kernel for the velocity $\dot{x}$ so that it exactly compensates for the resulting colored noise. Finally, from Eq.~\eqref{eq:2dsystem}, we obtain the following GLE for the dynamics of $x$ only \cite{supplemental_material}:
\begin{equation}
    \int_{-\infty}^{t} \mathrm{d}s ~\Gamma(t-s)\dot{x}_s = f^x_t - \mathcal{A}^{12}~x_t - \eta^x_t
    \label{GLE}
\end{equation}
where the resulting noise satisfies:
\begin{equation*}
    \langle \eta^x_{t} \eta^x_{t'} \rangle = 2 \delta(t-t') - \mathcal{A}^{12}~e^{A_{22}|t-t'|} = \Gamma(|t-t'|) \;,
\end{equation*}
and $\mathcal{A}^{12} = A_{12} A_{21}/A_{22}$ is a global parameter controlling the effective dynamics of the observed DOF, $x_t$. Notice that this approach prescribes the exact integration of the unobserved DOF, and it is not a coarse-graining procedure. As such, it preserves the dynamics of the system, by definition, independently of the thermodynamic properties. 
%even if the thermodynamics of the resulting equation is still not completely understood {\color{orange}{MC: substitute with "independently of the thermodynamic properties"}}. 
In this simple scenario, the entropy production of Eq.~\eqref{GLE} can be defined as before, $\sigma_{\rm tot} = \langle (f_t^x - \mathcal{A}^{12} x_t) \circ \dot{x}_t \rangle$, and coincides with Eq.~\eqref{diss} \cite{supplemental_material}. Hence, when the hidden DOF that acts as an effective bath is \textit{non-dissipative} - $y_t$ in this case - the resulting GLE provides a thermodynamically consistent description of the system and the entropy production rate can be calculated using the customary formula, Eq.~\eqref{sekimoto}. As we will discuss later, the crucial point is that Eq.~\eqref{GLE} has only memoryless forces.

Since the explicit integration of a \textit{non-dissipative} DOF leads to a GLE for the observed one, we can iterate the above procedure to take into account the effect of multiple \textit{non-dissipative} hidden variables. Consider $n$ observed DOFs, $x^i_t$, following $n$ GLEs when some hidden \textit{non-dissipative} DOFs have been already integrated out. They are coupled reciprocally to one additional hidden DOF, $y_t$: %\sout{that we would like to integrate out:}
\begin{equation}
\begin{split}
    \int_{-\infty}^{t} \mathrm{d}s \,\Gamma^{ij}(t-s)\dot{x}^j_s = f_t^{i} + A_{xy}^{i}y_t + \eta^{i}_t \\
    \dot{y}_t = A_{yy} y_t + A_{xy}^{i} x^i_t + \sqrt{2 D}\, \xi^y_t \, ,
\end{split}\label{MultiGamma}
\end{equation}
where Einstein's notation is used, $\langle \eta^i_{t} \eta^j_{t'} \rangle = \Gamma^{ij}(|t-t'|)$, and $\xi^y_t$ is a Gaussian white noise. Again, due to the reciprocal couplings between $y_t$ and $x^i_t$, the total entropy production rate coincides with the one of all $x^i_t$'s only, $\sigma_{\rm tot} = \langle F_t^i \circ \dot{x}^i_t \rangle$ (when $D=1$, for simplicity). By inserting $y_t$ into the evolution of $x_i$ once again, we can prove that Eq.~\eqref{MultiGamma} becomes a new GLE satisfying the second fluctuation dissipation theorem and driven by an effective force \cite{supplemental_material}:
\begin{equation}
    \int_{-\infty}^{t} \mathrm{d}s \,\widetilde{\Gamma}^{\,ij}(t-s)\dot{x}^i_s = f_t^{i}-\mathcal{A}^{ij}x^{j}_t+\widetilde{\eta}^{\,i}_t \, ,
    \label{GLEmulti}
\end{equation}
where $\mathcal{A}^{ij} = A^i_{xy} A^j_{xy}/A_{yy}$, and $\widetilde{\Gamma}$ is a modified memory kernel encapsulating the timescale of $y$:
\begin{equation*}
    \widetilde{\Gamma}^{\,ij}(|\tau|) = \Gamma^{\,ij}(|\tau|) - \mathcal{A}^{ij} e^{A_{yy}|\tau|}
\end{equation*}
and with $\langle \widetilde{\eta}^i_t \widetilde{\eta}^j_{t'} \rangle = \widetilde{\Gamma}^{\,ij}(|t-t'|)$. From Eq.~\eqref{GLE} and Eq.~\eqref{GLEmulti}, it is worth noting that the number of independent parameters in the effective dynamics of the observed DOFs reduces by one for each \textit{non-dissipative} DOF that is integrated out. This is clear as both the effective coupling and the noise kernel depend only on the global parameter $\mathcal{A}^{12}$.

By applying Eq.~\eqref{sekimoto} to Eq.~\eqref{GLEmulti}, the resulting entropy production rate in a NESS can be computed as follows \cite{supplemental_material}:
%By applying the Sekimoto's formula directly to Eq.~\eqref{GLEmulti}, the resulting entropy production rate at stationarity can be computed as follows \cite{supplemental_material}:
\begin{gather}
    \sigma_{\rm GLE}^{(x)} = \left\langle \left( f^i - \mathcal{A}^{ij}x_t^j \right) \circ \dot{x}^i_t \right\rangle = \\
    = \langle f_t^i \circ \dot{x}^i_t \rangle - \frac{\mathcal{A}^{ij}}{2} \left( \dotnice{C}^x_{ij}(0)-\dotnice{C}^x_{ji}(0) \right) = \langle f_t^i \circ \dot{x}^i_t \rangle \, \nonumber
\end{gather}
and coincides with the total entropy production rate $\sigma_{\rm tot}$. Here, we define $\dotnice{C}^x_{ij}(0)=\partial_t C_{x^i x^j}(t)|_{t=0}$. Note that the second term on the last line vanishes since it is a contraction between a symmetric and an anti-symmetric tensor. The same derivation can be carried out in the presence of non-linearities acting on the observed DOFs. Since the entropy production rate depends solely on the dynamics of observed DOFs, also their thermodynamics depend on a reduced set of parameters.

The first finding of this Letter is that hidden variables coupled reciprocally to observable DOFs and evolving on comparable timescales, i.e. \textit{non-dissipative}, naturally lead to an effective GLE description that is thermodynamically consistent. Indeed, the additional timescales are captured by the memory kernel and the estimated entropy production rate, using Eq.~\eqref{sekimoto}, coincides with the one of the whole system. Crucially, the presence of reciprocal couplings does not introduce any dissipative effect into the observed DOFs caused by the presence of hidden variables. As a consequence, we can summarize these statements claiming that hidden \textit{non-dissipative} DOFs characterize passive baths, i.e., environments that do not contribute to the total dissipation, and a description in terms of a GLE can be employed. An interesting additional observation is that the integration of a \textit{non-dissipative} DOF reduces by one the number of free parameters both in the system's dynamics and thermodynamics. Thus, the number of \textit{non-dissipative} DOFs determines the dimensionality of the isoentropic hypersurfaces in the parameter space.

We now go back the the simple scenario of two coupled DOFs, $x$ and $y$, described by Eq.~\eqref{eq:2dsystem}. We focus on the setting in which couplings are not symmetric, i.e., non-reciprocal, describing the case in which both variables are out of equilibrium and contribute to the total entropy production rate. We always consider $A_{11}$ and $A_{22}$ to be negative, and $f_t^x = f_t^y = 0$ for simplicity. The entropy production rate at stationarity is obtained as before:
\begin{gather}
    \sigma_{tot} = A^2_{12} C_{yy}(0) + A^2_{21} C_{xx}(0) ~+ \\
    + \left( A_{11} A_{12} + A_{22} A_{21} \right) C_{xy}(0) = - \frac{(A_{12} - A_{12})^2}{A_{11} + A_{22}} \nonumber
\end{gather}
where $C_{a b}(t-s) = \langle a_t b_s \rangle$ are the correlation functions. This is non-zero only when $A_{12} \neq A_{21}$, as expected.

As above, consider $x_t$ to be the only observed DOF. Since the couplings between $x_t$ and $y_t$ are non-reciprocal, we define $y_t$ as an \textit{dissipative} DOF and it contributes to the total dissipation of the system. By integrating out $y_t$ explicitly, i.e., solving its dynamics, and inserting the corresponding solution into the equation for $x_t$, one obtains the following closed equation for the $x_t$ only:
\begin{gather}\label{xDOF}
    \dot{x}_t = A_{11} x_t + \sqrt{2D} \,\xi^x_t + \\
    + ~A_{12} \int_{-\infty}^t {\rm d}s ~e^{-A_{22}(t-s)} \left( \sqrt{2D} \,\xi^y_s + A_{21} x_s \right) \nonumber
\end{gather}
It might not come as a surprise that, by evaluating the stationary entropy production rate from Eq.~\eqref{xDOF}, identifying as a stochastic force with memory all the terms that are different from $\sqrt{2D} \,\xi^x_t$, the intrinsic noise, we obtain \cite{supplemental_material}:
\begin{gather}
    \left\langle \left( \dot{x}_t - \sqrt{2}\,\xi^x_t \right) \circ \dot{x}_t \right\rangle = \langle F_t^x \circ \dot{x}_t \rangle = \nonumber \\
    = \frac{A_{12} \left( A_{21} - A_{12} \right)}{A_{11} + A_{22}} = \sigma_x
\end{gather}
with $D=1$ without loss of generality to simplify the notation. It coincides with the contribution to the stationary entropy production rate provided by $x$ alone. What is less obvious is that $\sigma_x$ can take negative values when $x$ absorbs energy from the environment, being only a part of $\sigma_{tot} > 0$. In such scenario, $\sigma_y$ has to be positive and greater than $-\sigma_x$. As a result of the exact integration of the unobserved DOF that acts as an effective bath, $x_t$ turns out to be a non-Markovian DOF. Ascribing the physical origin of this non-Markovianity to a stochastic force with memory, as above, results in a critical loss of information about the system's dissipation. However, the resulting sign of $\sigma_x$ is consistent with the role played by $x_t$ in determining the total entropy production rate, i.e., a dissipative ($\sigma_x > 0$) or a pumping ($\sigma_x < 0)$ DOF.

Since $y_t$ acts as an effective bath for $x_t$, one could construct, from Eq.~\eqref{xDOF}, a corresponding GLE using a procedure identical to the one employed before. Indeed, the two noise-dependent terms in Eq.~\eqref{xDOF} give rise to an effective colored noise, $\eta_t^x$, which is the same appearing in Eq.~\eqref{GLE}. Then, we reconstruct a dissipation kernel for $x_t$ that balances exactly the correlation of this noise. In this way, we formally identify a new non-Markovian bath, i.e., with memory, that satisfies by construction the fluctuation-dissipation theorem and results from the way $y_t$ influences the dynamics of $x_t$. The resulting GLE reads \cite{supplemental_material}:
\begin{equation}
    \int_{-\infty}^{t} \mathrm{d}s ~\Gamma(t-s)\dot{x}_s = \tilde{A}_{11} x_t + \int_{-\infty}^t \mathrm{d}s ~K_f(t-s) x_s + \eta^x_t \nonumber
\end{equation}
where $\tilde{A}_{11}$ is a modified coupling and $K_f$ a memory kernel proportional to $(A_{12} - A_{21})$. The explicit expressions of $\tilde{A}_{11}$ and $K_f$ are given in \cite{supplemental_material}. Crucially, integrating out \textit{dissipative} DOF does not reduce the number of free parameters as in the case of \textit{non-dissipative} DOFs. A striking difference with respect to Eq.~\eqref{GLE} is that an additional non-Markovian force naturally appears. As a consequence, the corresponding stationary entropy production rate reads \cite{supplemental_material}:
\begin{gather}
    \sigma_{GLE}^{(x)} = \left\langle \left( \int_{-\infty}^t \mathrm{d}s ~K_f(t-s) x_s \right) \circ \dot{x}_t \right\rangle = \\
    = \sigma_x \left( 1 + \frac{\alpha^{-1} \mathcal{A}^{12}}{2 \left(A_{11} + A_{22}\right) - \mathcal{A}^{12}} \right) \nonumber
\end{gather}
where $\alpha = A_{21}/A_{12}$ quantifies the non-reciprocity of couplings. Manifestly, this entropy production rate is different from both $\sigma_{tot}$ and $\sigma_x$, and it can also take negative values. As such, it does not contain any useful thermodynamic information on the system. We ascribe this discrepancy to the presence of non-reciprocal interactions between $x_t$ and $y_t$ that cause the existence of this additional force with memory. In other words, trying to reconstruct an effective bath integrating out \textit{non-dissipative} DOFs leads to thermodynamic inconsistency and errors in estimating the system's dissipation. These problems arise for two reasons: i) $y_t$ is out of equilibrium ($\sigma_y \neq 0$) and its dissipative contribution cannot be captured by a GLE; ii) $y_t$ is at equilibrium ($\sigma_y = 0$) but drives the dynamics of $x$ in a non-reciprocal way, as for the case of the Active Ornstein-Uhlenbeck (AOU) process \cite{szamel2014self,sevilla2019generalized}. In both scenarios, the action of \textit{non-dissipative} unobservable Markovian DOFs modifies the dissipation of the whole system and, as such, it has to be associated with active baths and cannot be consistently described using an effective GLE.

\begin{figure}[t]
    \centering
    \includegraphics[width=\columnwidth]{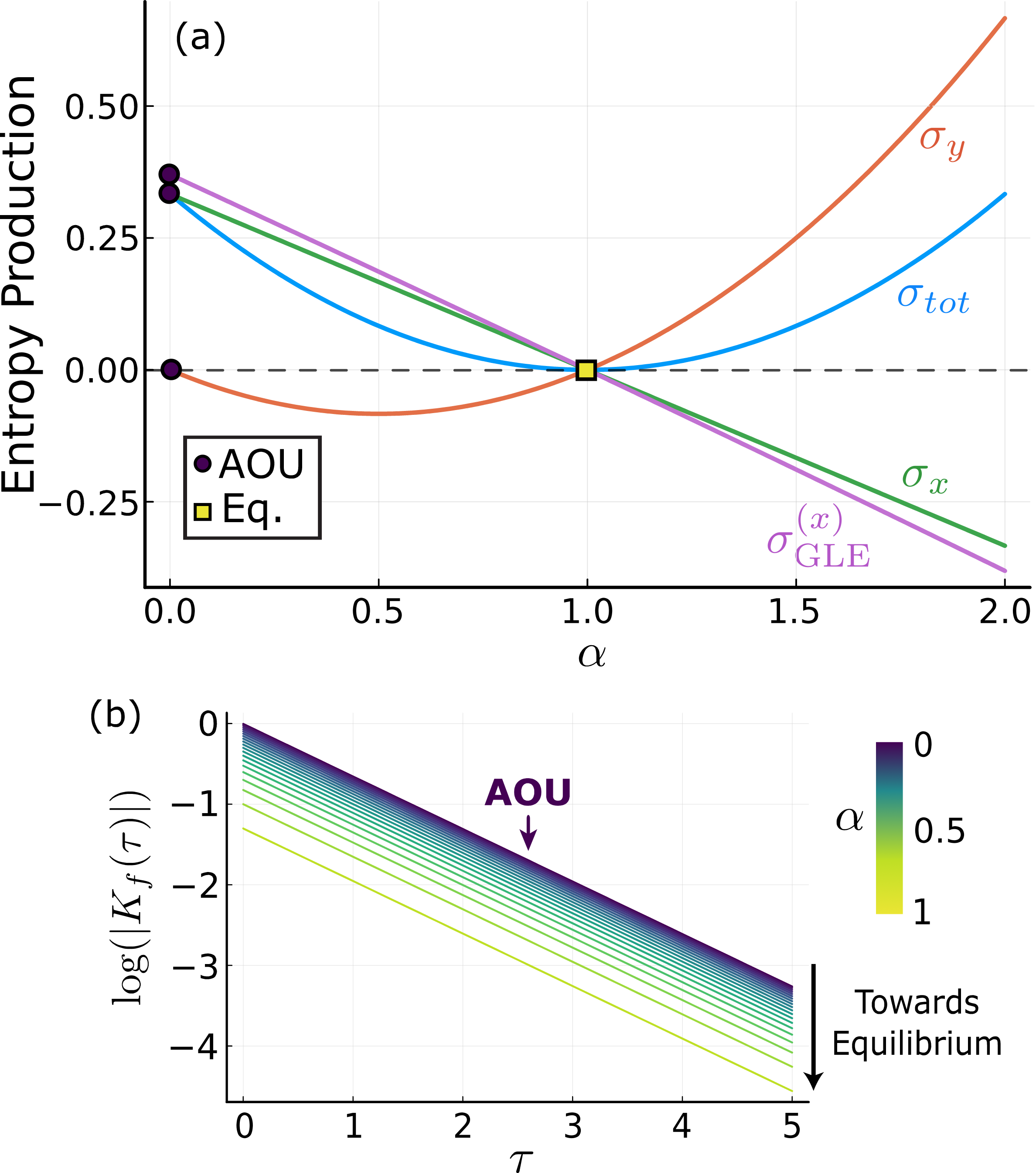}
    \caption{Dissipation and memory in a $2D$ system. (a) Different contributions to the stationary entropy production rate for increasing $\alpha$ ($\alpha = 0$, AOU process; $\alpha = 1$, passive bath). While $\sigma_{tot}$ is exact and non-negative, all the other terms have no definite sign. (b) The logarithm of the force kernel is shown as a function of $\tau$ for different values of $\alpha$ (in color scale). As $\alpha$ approaches $1$, non-Markovian effects disappears. Details are reported in the main text.}
    \label{fig1}
\end{figure}

We now show a clarifying example to visualize our results. Consider the system in Eq.~\eqref{eq:2dsystem}, with $A_{12} = D = 1 = - A_{ii}$ for $i = 1, 2$. For $\alpha = 0$ (i.e., $A_{21} = 0$), this model is an AOU process since $y_t$ follows an independent Ornstein-Uhlenbeck dynamics. In this case, $\sigma_y = 0$, but $\sigma_{GLE}^{(x)} \neq \sigma_x$. For $\alpha = 1$, the system restores reciprocal interactions and $\sigma_{GLE}^{(x)} = \sigma_x$. In Fig.~\ref{fig1}a, we show all the different entropic contributions for increasing values of $\alpha$, showing that $\sigma_{GLE}^{(x)}$ can either under- or over-estimate both $\sigma_{tot}$ and $\sigma_x$ depending on the parameter. Moreover, only $\sigma_{tot} \geq 0$, vanishing at equilibrium. In Fig.~\ref{fig1}b, we plot the force kernel $K_f$ as a function of $\tau \equiv (t-s)$ for different values of $\alpha$, highlighting the presence of a non-Markovian additional force in the resulting GLE when $y$ is a \textit{non-dissipative} DOF ($\alpha \neq 1$).

To wrap up, in this Letter, we identify two classes of stochastic variables, i.e., \textit{non-dissipative} and \textit{dissipative}. The first ones characterize DOFs that are coupled reciprocally to the observed variables. As such, they do not contribute to the total dissipation of the system and determine the dimensionality of isoentropic hypersurface in the parameter space. The second ones are DOFs coupled non-reciprocally to the observable (e.g., tracer particle), even if they do not necessarily contribute to the total entropy production rate, as for AOU process \cite{supplemental_material}. 

This classification appears to be crucial when discussing effective models that rely on the integration of hidden variables. Indeed, we show that, when these hidden DOFs are \textit{non-dissipative}, a description in terms of a GLE is allowed and thermodynamically consistent. Thus, the identification of a memory kernel for the effective bath does not affect the thermodynamics of the whole system. As such, we state that \textit{non-dissipative} DOFs should be employed only when dealing with passive baths. Vice-versa, when a bath is passive but evolves on a finite time-scale, any compatible microscopic Markovian model should only encompass \textit{non-dissipative} DOFs.

On the contrary, when hidden DOFs are \textit{dissipative}, GLEs are proven to be thermodynamically inconsistent. Furthermore, they do not provide any useful information on dissipation, by either under- or over-estimating the total entropy production rate of the whole system. This result is unexpected and apparently in contrast with the usual statement that entropy production decreases with coarse-graining procedures \cite{esposito2012stochastic,busiello2019entropy}. In fact, the framework employed here prescribes an exact integration which is substantially different from standard coarse-graining approaches. We conclude that the introduction of a memory kernel to describe the effective bath modifies the system's thermodynamics. As such, \textit{dissipative} DOFs have to be associated to microscopic Markovian models for active baths, and their precise identification is instrumental for the correct evaluation of the entropy production rate.

On this particular aspect, we are aware of the existence of previous works that discussed how to estimate the entropy production rate of active systems \cite{zamponi2005fluctuation,mandal2017entropy,pietzonka2017entropy,caprini2019entropy,dabelow2019irreversibility}. At variance with microscopic models, when dealing with mesoscopic dynamics where a certain degree of coarse-graining has already incurred, the connection between dynamics and thermodynamics is affected by the amount of accessible information on all DOFs. For example, when defining the entropy production as irreversibility, one has to assume the behavior of the active noise with respect to time inversion, which in turn depends on the variables that generate it \cite{dabelow2019irreversibility}. As a matter of fact, as memory usually hinders the presence of unaccessible DOFs, the correct estimation of thermodynamic features boils down to unravel the underlying Markovian dynamics where thermal white noise is the only source of stochasticity. Given these premises, this work sheds some light on limitations and potentialities of different (dynamically equivalent) models in estimating the entropy production rate of the system from a subset of observed variables. The fundamental results highlighted in this Letter pave the way for a more profound understanding of how an active bath can be fully characterized and, most importantly, how they can be integrated out preserving at best all thermodynamic properties.

\section{Acknowledgments}

The authors thank F. J\"ulicher for the insightful discussions that stimulated this study, and M. Baiesi for careful reading and useful comments.

%\bibliography{Bibliography}

\begin{thebibliography}{37}%
\makeatletter
\providecommand \@ifxundefined [1]{%
 \@ifx{#1\undefined}
}%
\providecommand \@ifnum [1]{%
 \ifnum #1\expandafter \@firstoftwo
 \else \expandafter \@secondoftwo
 \fi
}%
\providecommand \@ifx [1]{%
 \ifx #1\expandafter \@firstoftwo
 \else \expandafter \@secondoftwo
 \fi
}%
\providecommand \natexlab [1]{#1}%
\providecommand \enquote  [1]{``#1''}%
\providecommand \bibnamefont  [1]{#1}%
\providecommand \bibfnamefont [1]{#1}%
\providecommand \citenamefont [1]{#1}%
\providecommand \href@noop [0]{\@secondoftwo}%
\providecommand \href [0]{\begingroup \@sanitize@url \@href}%
\providecommand \@href[1]{\@@startlink{#1}\@@href}%
\providecommand \@@href[1]{\endgroup#1\@@endlink}%
\providecommand \@sanitize@url [0]{\catcode `\\12\catcode `\$12\catcode
  `\&12\catcode `\#12\catcode `\^12\catcode `\_12\catcode `\%12\relax}%
\providecommand \@@startlink[1]{}%
\providecommand \@@endlink[0]{}%
\providecommand \url  [0]{\begingroup\@sanitize@url \@url }%
\providecommand \@url [1]{\endgroup\@href {#1}{\urlprefix }}%
\providecommand \urlprefix  [0]{URL }%
\providecommand \Eprint [0]{\href }%
\providecommand \doibase [0]{https://doi.org/}%
\providecommand \selectlanguage [0]{\@gobble}%
\providecommand \bibinfo  [0]{\@secondoftwo}%
\providecommand \bibfield  [0]{\@secondoftwo}%
\providecommand \translation [1]{[#1]}%
\providecommand \BibitemOpen [0]{}%
\providecommand \bibitemStop [0]{}%
\providecommand \bibitemNoStop [0]{.\EOS\space}%
\providecommand \EOS [0]{\spacefactor3000\relax}%
\providecommand \BibitemShut  [1]{\csname bibitem#1\endcsname}%
\let\auto@bib@innerbib\@empty
%</preamble>
\bibitem [{\citenamefont {Fodor}\ \emph {et~al.}(2022)\citenamefont {Fodor},
  \citenamefont {Jack},\ and\ \citenamefont
  {Cates}}]{fodor2022irreversibility}%
  \BibitemOpen
  \bibfield  {author} {\bibinfo {author} {\bibfnamefont {{\'E}.}~\bibnamefont
  {Fodor}}, \bibinfo {author} {\bibfnamefont {R.~L.}\ \bibnamefont {Jack}},\
  and\ \bibinfo {author} {\bibfnamefont {M.~E.}\ \bibnamefont {Cates}},\
  }\bibfield  {title} {\bibinfo {title} {Irreversibility and biased ensembles
  in active matter: Insights from stochastic thermodynamics},\ }\href@noop {}
  {\bibfield  {journal} {\bibinfo  {journal} {Annual Review of Condensed Matter
  Physics}\ }\textbf {\bibinfo {volume} {13}},\ \bibinfo {pages} {215}
  (\bibinfo {year} {2022})}\BibitemShut {NoStop}%
\bibitem [{\citenamefont {J{\"u}licher}\ \emph {et~al.}(2018)\citenamefont
  {J{\"u}licher}, \citenamefont {Grill},\ and\ \citenamefont
  {Salbreux}}]{julicher2018hydrodynamic}%
  \BibitemOpen
  \bibfield  {author} {\bibinfo {author} {\bibfnamefont {F.}~\bibnamefont
  {J{\"u}licher}}, \bibinfo {author} {\bibfnamefont {S.~W.}\ \bibnamefont
  {Grill}},\ and\ \bibinfo {author} {\bibfnamefont {G.}~\bibnamefont
  {Salbreux}},\ }\bibfield  {title} {\bibinfo {title} {Hydrodynamic theory of
  active matter},\ }\href@noop {} {\bibfield  {journal} {\bibinfo  {journal}
  {Reports on Progress in Physics}\ }\textbf {\bibinfo {volume} {81}},\
  \bibinfo {pages} {076601} (\bibinfo {year} {2018})}\BibitemShut {NoStop}%
\bibitem [{\citenamefont {Elgeti}\ \emph {et~al.}(2015)\citenamefont {Elgeti},
  \citenamefont {Winkler},\ and\ \citenamefont {Gompper}}]{elgeti2015physics}%
  \BibitemOpen
  \bibfield  {author} {\bibinfo {author} {\bibfnamefont {J.}~\bibnamefont
  {Elgeti}}, \bibinfo {author} {\bibfnamefont {R.~G.}\ \bibnamefont
  {Winkler}},\ and\ \bibinfo {author} {\bibfnamefont {G.}~\bibnamefont
  {Gompper}},\ }\bibfield  {title} {\bibinfo {title} {Physics of
  microswimmers—single particle motion and collective behavior: a review},\
  }\href@noop {} {\bibfield  {journal} {\bibinfo  {journal} {Reports on
  progress in physics}\ }\textbf {\bibinfo {volume} {78}},\ \bibinfo {pages}
  {056601} (\bibinfo {year} {2015})}\BibitemShut {NoStop}%
\bibitem [{\citenamefont {Palacci}\ \emph {et~al.}(2013)\citenamefont
  {Palacci}, \citenamefont {Sacanna}, \citenamefont {Steinberg}, \citenamefont
  {Pine},\ and\ \citenamefont {Chaikin}}]{palacci2013living}%
  \BibitemOpen
  \bibfield  {author} {\bibinfo {author} {\bibfnamefont {J.}~\bibnamefont
  {Palacci}}, \bibinfo {author} {\bibfnamefont {S.}~\bibnamefont {Sacanna}},
  \bibinfo {author} {\bibfnamefont {A.~P.}\ \bibnamefont {Steinberg}}, \bibinfo
  {author} {\bibfnamefont {D.~J.}\ \bibnamefont {Pine}},\ and\ \bibinfo
  {author} {\bibfnamefont {P.~M.}\ \bibnamefont {Chaikin}},\ }\bibfield
  {title} {\bibinfo {title} {Living crystals of light-activated colloidal
  surfers},\ }\href@noop {} {\bibfield  {journal} {\bibinfo  {journal}
  {Science}\ }\textbf {\bibinfo {volume} {339}},\ \bibinfo {pages} {936}
  (\bibinfo {year} {2013})}\BibitemShut {NoStop}%
\bibitem [{\citenamefont {Ghosh}\ \emph {et~al.}(2021)\citenamefont {Ghosh},
  \citenamefont {Somasundar},\ and\ \citenamefont {Sen}}]{ghosh2021enzymes}%
  \BibitemOpen
  \bibfield  {author} {\bibinfo {author} {\bibfnamefont {S.}~\bibnamefont
  {Ghosh}}, \bibinfo {author} {\bibfnamefont {A.}~\bibnamefont {Somasundar}},\
  and\ \bibinfo {author} {\bibfnamefont {A.}~\bibnamefont {Sen}},\ }\bibfield
  {title} {\bibinfo {title} {Enzymes as active matter},\ }\href@noop {}
  {\bibfield  {journal} {\bibinfo  {journal} {Annual Review of Condensed Matter
  Physics}\ }\textbf {\bibinfo {volume} {12}},\ \bibinfo {pages} {177}
  (\bibinfo {year} {2021})}\BibitemShut {NoStop}%
\bibitem [{\citenamefont {Granek}\ \emph {et~al.}(2022)\citenamefont {Granek},
  \citenamefont {Kafri},\ and\ \citenamefont {Tailleur}}]{granek2022anomalous}%
  \BibitemOpen
  \bibfield  {author} {\bibinfo {author} {\bibfnamefont {O.}~\bibnamefont
  {Granek}}, \bibinfo {author} {\bibfnamefont {Y.}~\bibnamefont {Kafri}},\ and\
  \bibinfo {author} {\bibfnamefont {J.}~\bibnamefont {Tailleur}},\ }\bibfield
  {title} {\bibinfo {title} {Anomalous transport of tracers in active baths},\
  }\href@noop {} {\bibfield  {journal} {\bibinfo  {journal} {Physical Review
  Letters}\ }\textbf {\bibinfo {volume} {129}},\ \bibinfo {pages} {038001}
  (\bibinfo {year} {2022})}\BibitemShut {NoStop}%
\bibitem [{\citenamefont {Liang}\ \emph {et~al.}(2022)\citenamefont {Liang},
  \citenamefont {Busiello},\ and\ \citenamefont
  {De~Los~Rios}}]{liang2022emergent}%
  \BibitemOpen
  \bibfield  {author} {\bibinfo {author} {\bibfnamefont {S.}~\bibnamefont
  {Liang}}, \bibinfo {author} {\bibfnamefont {D.~M.}\ \bibnamefont
  {Busiello}},\ and\ \bibinfo {author} {\bibfnamefont {P.}~\bibnamefont
  {De~Los~Rios}},\ }\bibfield  {title} {\bibinfo {title} {Emergent
  thermophoretic behavior in chemical reaction systems},\ }\href@noop {}
  {\bibfield  {journal} {\bibinfo  {journal} {New Journal of Physics}\ }\textbf
  {\bibinfo {volume} {24}},\ \bibinfo {pages} {123006} (\bibinfo {year}
  {2022})}\BibitemShut {NoStop}%
\bibitem [{\citenamefont {Jin}\ and\ \citenamefont
  {Zhang}(2021)}]{jin2021collective}%
  \BibitemOpen
  \bibfield  {author} {\bibinfo {author} {\bibfnamefont {D.}~\bibnamefont
  {Jin}}\ and\ \bibinfo {author} {\bibfnamefont {L.}~\bibnamefont {Zhang}},\
  }\bibfield  {title} {\bibinfo {title} {Collective behaviors of magnetic
  active matter: recent progress toward reconfigurable, adaptive, and
  multifunctional swarming micro/nanorobots},\ }\href@noop {} {\bibfield
  {journal} {\bibinfo  {journal} {Accounts of Chemical Research}\ }\textbf
  {\bibinfo {volume} {55}},\ \bibinfo {pages} {98} (\bibinfo {year}
  {2021})}\BibitemShut {NoStop}%
\bibitem [{\citenamefont {Bois}\ \emph {et~al.}(2011)\citenamefont {Bois},
  \citenamefont {J{\"u}licher},\ and\ \citenamefont {Grill}}]{bois2011pattern}%
  \BibitemOpen
  \bibfield  {author} {\bibinfo {author} {\bibfnamefont {J.~S.}\ \bibnamefont
  {Bois}}, \bibinfo {author} {\bibfnamefont {F.}~\bibnamefont {J{\"u}licher}},\
  and\ \bibinfo {author} {\bibfnamefont {S.~W.}\ \bibnamefont {Grill}},\
  }\bibfield  {title} {\bibinfo {title} {Pattern formation in active fluids},\
  }\href@noop {} {\bibfield  {journal} {\bibinfo  {journal} {Biophysical
  Journal}\ }\textbf {\bibinfo {volume} {100}},\ \bibinfo {pages} {445a}
  (\bibinfo {year} {2011})}\BibitemShut {NoStop}%
%\bibitem [{\citenamefont {Liebchen}\ and\ \citenamefont
%  {Levis}(2017)}]{liebchen2017collective}%
%  \BibitemOpen
%  \bibfield  {author} {\bibinfo {author} {\bibfnamefont {B.}~\bibnamefont
%  {Liebchen}}\ and\ \bibinfo {author} {\bibfnamefont {D.}~\bibnamefont
%  {Levis}},\ }\bibfield  {title} {\bibinfo {title} {Collective behavior of
%  chiral active matter: Pattern formation and enhanced flocking},\ }\href@noop
%  {} {\bibfield  {journal} {\bibinfo  {journal} {Physical review letters}\
%  }\textbf {\bibinfo {volume} {119}},\ \bibinfo {pages} {058002} (\bibinfo
%  {year} {2017})}\BibitemShut {NoStop}%
\bibitem [{\citenamefont {Cates}\ and\ \citenamefont
  {Tailleur}(2015)}]{cates2015motility}%
  \BibitemOpen
  \bibfield  {author} {\bibinfo {author} {\bibfnamefont {M.~E.}\ \bibnamefont
  {Cates}}\ and\ \bibinfo {author} {\bibfnamefont {J.}~\bibnamefont
  {Tailleur}},\ }\bibfield  {title} {\bibinfo {title} {Motility-induced phase
  separation},\ }\href@noop {} {\bibfield  {journal} {\bibinfo  {journal}
  {Annu. Rev. Condens. Matter Phys.}\ }\textbf {\bibinfo {volume} {6}},\
  \bibinfo {pages} {219} (\bibinfo {year} {2015})}\BibitemShut {NoStop}%
\bibitem [{\citenamefont {Maes}(2020)}]{maes2020fluctuating}%
  \BibitemOpen
  \bibfield  {author} {\bibinfo {author} {\bibfnamefont {C.}~\bibnamefont
  {Maes}},\ }\bibfield  {title} {\bibinfo {title} {Fluctuating motion in an
  active environment},\ }\href@noop {} {\bibfield  {journal} {\bibinfo
  {journal} {Physical Review Letters}\ }\textbf {\bibinfo {volume} {125}},\
  \bibinfo {pages} {208001} (\bibinfo {year} {2020})}\BibitemShut {NoStop}%
\bibitem [{\citenamefont {Pietzonka}\ and\ \citenamefont
  {Seifert}(2017)}]{pietzonka2017entropy}%
  \BibitemOpen
  \bibfield  {author} {\bibinfo {author} {\bibfnamefont {P.}~\bibnamefont
  {Pietzonka}}\ and\ \bibinfo {author} {\bibfnamefont {U.}~\bibnamefont
  {Seifert}},\ }\bibfield  {title} {\bibinfo {title} {Entropy production of
  active particles and for particles in active baths},\ }\href@noop {}
  {\bibfield  {journal} {\bibinfo  {journal} {Journal of Physics A:
  Mathematical and Theoretical}\ }\textbf {\bibinfo {volume} {51}},\ \bibinfo
  {pages} {01LT01} (\bibinfo {year} {2017})}\BibitemShut {NoStop}%
\bibitem [{\citenamefont {Banerjee}\ \emph {et~al.}(2022)\citenamefont
  {Banerjee}, \citenamefont {Jack},\ and\ \citenamefont
  {Cates}}]{banerjee2022tracer}%
  \BibitemOpen
  \bibfield  {author} {\bibinfo {author} {\bibfnamefont {T.}~\bibnamefont
  {Banerjee}}, \bibinfo {author} {\bibfnamefont {R.~L.}\ \bibnamefont {Jack}},\
  and\ \bibinfo {author} {\bibfnamefont {M.~E.}\ \bibnamefont {Cates}},\
  }\bibfield  {title} {\bibinfo {title} {Tracer dynamics in one dimensional
  gases of active or passive particles},\ }\href@noop {} {\bibfield  {journal}
  {\bibinfo  {journal} {Journal of Statistical Mechanics: Theory and
  Experiment}\ }\textbf {\bibinfo {volume} {2022}},\ \bibinfo {pages} {013209}
  (\bibinfo {year} {2022})}\BibitemShut {NoStop}%
\bibitem [{\citenamefont {Caspi}\ \emph {et~al.}(2000)\citenamefont {Caspi},
  \citenamefont {Granek},\ and\ \citenamefont {Elbaum}}]{caspi2000enhanced}%
  \BibitemOpen
  \bibfield  {author} {\bibinfo {author} {\bibfnamefont {A.}~\bibnamefont
  {Caspi}}, \bibinfo {author} {\bibfnamefont {R.}~\bibnamefont {Granek}},\ and\
  \bibinfo {author} {\bibfnamefont {M.}~\bibnamefont {Elbaum}},\ }\bibfield
  {title} {\bibinfo {title} {Enhanced diffusion in active intracellular
  transport},\ }\href@noop {} {\bibfield  {journal} {\bibinfo  {journal}
  {Physical Review Letters}\ }\textbf {\bibinfo {volume} {85}},\ \bibinfo
  {pages} {5655} (\bibinfo {year} {2000})}\BibitemShut {NoStop}%
\bibitem [{\citenamefont {Xu}\ \emph {et~al.}(2019)\citenamefont {Xu},
  \citenamefont {Ross}, \citenamefont {Valdez},\ and\ \citenamefont
  {Sen}}]{xu2019direct}%
  \BibitemOpen
  \bibfield  {author} {\bibinfo {author} {\bibfnamefont {M.}~\bibnamefont
  {Xu}}, \bibinfo {author} {\bibfnamefont {J.~L.}\ \bibnamefont {Ross}},
  \bibinfo {author} {\bibfnamefont {L.}~\bibnamefont {Valdez}},\ and\ \bibinfo
  {author} {\bibfnamefont {A.}~\bibnamefont {Sen}},\ }\bibfield  {title}
  {\bibinfo {title} {Direct single molecule imaging of enhanced enzyme
  diffusion},\ }\href@noop {} {\bibfield  {journal} {\bibinfo  {journal}
  {Physical review letters}\ }\textbf {\bibinfo {volume} {123}},\ \bibinfo
  {pages} {128101} (\bibinfo {year} {2019})}\BibitemShut {NoStop}%
\bibitem [{\citenamefont {Mori}(1965)}]{mori1965transport}%
  \BibitemOpen
  \bibfield  {author} {\bibinfo {author} {\bibfnamefont {H.}~\bibnamefont
  {Mori}},\ }\bibfield  {title} {\bibinfo {title} {Transport, collective
  motion, and brownian motion},\ }\href@noop {} {\bibfield  {journal} {\bibinfo
   {journal} {Progress of theoretical physics}\ }\textbf {\bibinfo {volume}
  {33}},\ \bibinfo {pages} {423} (\bibinfo {year} {1965})}\BibitemShut
  {NoStop}%
\bibitem [{\citenamefont {Kubo}(1966)}]{kubo1966fluctuation}%
  \BibitemOpen
  \bibfield  {author} {\bibinfo {author} {\bibfnamefont {R.}~\bibnamefont
  {Kubo}},\ }\bibfield  {title} {\bibinfo {title} {The fluctuation-dissipation
  theorem},\ }\href@noop {} {\bibfield  {journal} {\bibinfo  {journal} {Reports
  on progress in physics}\ }\textbf {\bibinfo {volume} {29}},\ \bibinfo {pages}
  {255} (\bibinfo {year} {1966})}\BibitemShut {NoStop}%
\bibitem [{\citenamefont {Mai}\ and\ \citenamefont
  {Dhar}(2007)}]{mai2007nonequilibrium}%
  \BibitemOpen
  \bibfield  {author} {\bibinfo {author} {\bibfnamefont {T.}~\bibnamefont
  {Mai}}\ and\ \bibinfo {author} {\bibfnamefont {A.}~\bibnamefont {Dhar}},\
  }\bibfield  {title} {\bibinfo {title} {Nonequilibrium work fluctuations for
  oscillators in non-markovian baths},\ }\href@noop {} {\bibfield  {journal}
  {\bibinfo  {journal} {Physical Review E}\ }\textbf {\bibinfo {volume} {75}},\
  \bibinfo {pages} {061101} (\bibinfo {year} {2007})}\BibitemShut {NoStop}%
\bibitem [{\citenamefont {Di~Terlizzi}\ \emph {et~al.}(2020)\citenamefont
  {Di~Terlizzi}, \citenamefont {Ritort},\ and\ \citenamefont
  {Baiesi}}]{OurGLE}%
  \BibitemOpen
  \bibfield  {author} {\bibinfo {author} {\bibfnamefont {I.}~\bibnamefont
  {Di~Terlizzi}}, \bibinfo {author} {\bibfnamefont {F.}~\bibnamefont
  {Ritort}},\ and\ \bibinfo {author} {\bibfnamefont {M.}~\bibnamefont
  {Baiesi}},\ }\bibfield  {title} {\bibinfo {title} {{Explicit solution of the
  generalised {L}angevin equation}},\ }\href@noop {} {\bibfield  {journal}
  {\bibinfo  {journal} {J. Stat. Phys.}\ }\textbf {\bibinfo {volume} {181}},\
  \bibinfo {pages} {1609} (\bibinfo {year} {{2020}})}\BibitemShut {NoStop}%
\bibitem [{\citenamefont {Di~Terlizzi}\ and\ \citenamefont
  {Baiesi}(2020)}]{di2020thermodynamic}%
  \BibitemOpen
  \bibfield  {author} {\bibinfo {author} {\bibfnamefont {I.}~\bibnamefont
  {Di~Terlizzi}}\ and\ \bibinfo {author} {\bibfnamefont {M.}~\bibnamefont
  {Baiesi}},\ }\bibfield  {title} {\bibinfo {title} {A thermodynamic
  uncertainty relation for a system with memory},\ }\href@noop {} {\bibfield
  {journal} {\bibinfo  {journal} {Journal of Physics A: Mathematical and
  Theoretical}\ }\textbf {\bibinfo {volume} {53}},\ \bibinfo {pages} {474002}
  (\bibinfo {year} {2020})}\BibitemShut {NoStop}%
\bibitem [{\citenamefont {Sekimoto}(1998)}]{sekimoto1998langevin}%
  \BibitemOpen
  \bibfield  {author} {\bibinfo {author} {\bibfnamefont {K.}~\bibnamefont
  {Sekimoto}},\ }\bibfield  {title} {\bibinfo {title} {Langevin equation and
  thermodynamics},\ }\href@noop {} {\bibfield  {journal} {\bibinfo  {journal}
  {Progress of Theoretical Physics Supplement}\ }\textbf {\bibinfo {volume}
  {130}},\ \bibinfo {pages} {17} (\bibinfo {year} {1998})}\BibitemShut
  {NoStop}%
\bibitem [{\citenamefont {Esposito}(2012)}]{esposito2012stochastic}%
  \BibitemOpen
  \bibfield  {author} {\bibinfo {author} {\bibfnamefont {M.}~\bibnamefont
  {Esposito}},\ }\bibfield  {title} {\bibinfo {title} {Stochastic
  thermodynamics under coarse graining},\ }\href@noop {} {\bibfield  {journal}
  {\bibinfo  {journal} {Physical Review E}\ }\textbf {\bibinfo {volume} {85}},\
  \bibinfo {pages} {041125} (\bibinfo {year} {2012})}\BibitemShut {NoStop}%
\bibitem [{\citenamefont {Busiello}\ \emph {et~al.}(2019)\citenamefont
  {Busiello}, \citenamefont {Hidalgo},\ and\ \citenamefont
  {Maritan}}]{busiello2019entropy}%
  \BibitemOpen
  \bibfield  {author} {\bibinfo {author} {\bibfnamefont {D.~M.}\ \bibnamefont
  {Busiello}}, \bibinfo {author} {\bibfnamefont {J.}~\bibnamefont {Hidalgo}},\
  and\ \bibinfo {author} {\bibfnamefont {A.}~\bibnamefont {Maritan}},\
  }\bibfield  {title} {\bibinfo {title} {Entropy production for coarse-grained
  dynamics},\ }\href@noop {} {\bibfield  {journal} {\bibinfo  {journal} {New
  Journal of Physics}\ }\textbf {\bibinfo {volume} {21}},\ \bibinfo {pages}
  {073004} (\bibinfo {year} {2019})}\BibitemShut {NoStop}%
\bibitem [{\citenamefont {Yu}\ \emph {et~al.}(2021)\citenamefont {Yu},
  \citenamefont {Zhang},\ and\ \citenamefont {Tu}}]{yu2021inverse}%
  \BibitemOpen
  \bibfield  {author} {\bibinfo {author} {\bibfnamefont {Q.}~\bibnamefont
  {Yu}}, \bibinfo {author} {\bibfnamefont {D.}~\bibnamefont {Zhang}},\ and\
  \bibinfo {author} {\bibfnamefont {Y.}~\bibnamefont {Tu}},\ }\bibfield
  {title} {\bibinfo {title} {Inverse power law scaling of energy dissipation
  rate in nonequilibrium reaction networks},\ }\href@noop {} {\bibfield
  {journal} {\bibinfo  {journal} {Physical review letters}\ }\textbf {\bibinfo
  {volume} {126}},\ \bibinfo {pages} {080601} (\bibinfo {year}
  {2021})}\BibitemShut {NoStop}%
\bibitem [{\citenamefont {Busiello}\ \emph {et~al.}(2020)\citenamefont
  {Busiello}, \citenamefont {Gupta},\ and\ \citenamefont
  {Maritan}}]{busiello2020coarse}%
  \BibitemOpen
  \bibfield  {author} {\bibinfo {author} {\bibfnamefont {D.~M.}\ \bibnamefont
  {Busiello}}, \bibinfo {author} {\bibfnamefont {D.}~\bibnamefont {Gupta}},\
  and\ \bibinfo {author} {\bibfnamefont {A.}~\bibnamefont {Maritan}},\
  }\bibfield  {title} {\bibinfo {title} {Coarse-grained entropy production with
  multiple reservoirs: Unraveling the role of time scales and detailed balance
  in biology-inspired systems},\ }\href@noop {} {\bibfield  {journal} {\bibinfo
   {journal} {Physical Review Research}\ }\textbf {\bibinfo {volume} {2}},\
  \bibinfo {pages} {043257} (\bibinfo {year} {2020})}\BibitemShut {NoStop}%
\bibitem [{\citenamefont {Busiello}\ and\ \citenamefont
  {Maritan}(2019)}]{busiello2019entropy2}%
  \BibitemOpen
  \bibfield  {author} {\bibinfo {author} {\bibfnamefont {D.~M.}\ \bibnamefont
  {Busiello}}\ and\ \bibinfo {author} {\bibfnamefont {A.}~\bibnamefont
  {Maritan}},\ }\bibfield  {title} {\bibinfo {title} {Entropy production in
  master equations and fokker--planck equations: facing the coarse-graining and
  recovering the information loss},\ }\href@noop {} {\bibfield  {journal}
  {\bibinfo  {journal} {Journal of Statistical Mechanics: Theory and
  Experiment}\ }\textbf {\bibinfo {volume} {2019}},\ \bibinfo {pages} {104013}
  (\bibinfo {year} {2019})}\BibitemShut {NoStop}%
\bibitem [{\citenamefont {Ghosal}\ and\ \citenamefont
  {Bisker}(2022)}]{ghosal2022inferring}%
  \BibitemOpen
  \bibfield  {author} {\bibinfo {author} {\bibfnamefont {A.}~\bibnamefont
  {Ghosal}}\ and\ \bibinfo {author} {\bibfnamefont {G.}~\bibnamefont
  {Bisker}},\ }\bibfield  {title} {\bibinfo {title} {Inferring entropy
  production rate from partially observed langevin dynamics under
  coarse-graining},\ }\href@noop {} {\bibfield  {journal} {\bibinfo  {journal}
  {Physical Chemistry Chemical Physics}\ }\textbf {\bibinfo {volume} {24}},\
  \bibinfo {pages} {24021} (\bibinfo {year} {2022})}\BibitemShut {NoStop}%
\bibitem [{\citenamefont {Filliger}\ and\ \citenamefont
  {Reimann}(2007)}]{filliger2007brownian}%
  \BibitemOpen
  \bibfield  {author} {\bibinfo {author} {\bibfnamefont {R.}~\bibnamefont
  {Filliger}}\ and\ \bibinfo {author} {\bibfnamefont {P.}~\bibnamefont
  {Reimann}},\ }\bibfield  {title} {\bibinfo {title} {Brownian gyrator: A
  minimal heat engine on the nanoscale},\ }\href@noop {} {\bibfield  {journal}
  {\bibinfo  {journal} {Physical review letters}\ }\textbf {\bibinfo {volume}
  {99}},\ \bibinfo {pages} {230602} (\bibinfo {year} {2007})}\BibitemShut
  {NoStop}%
\bibitem [{\citenamefont {Young}\ \emph {et~al.}(2020)\citenamefont {Young},
  \citenamefont {Gorshkov}, \citenamefont {Foss-Feig},\ and\ \citenamefont
  {Maghrebi}}]{young2020nonequilibrium}%
  \BibitemOpen
  \bibfield  {author} {\bibinfo {author} {\bibfnamefont {J.~T.}\ \bibnamefont
  {Young}}, \bibinfo {author} {\bibfnamefont {A.~V.}\ \bibnamefont {Gorshkov}},
  \bibinfo {author} {\bibfnamefont {M.}~\bibnamefont {Foss-Feig}},\ and\
  \bibinfo {author} {\bibfnamefont {M.~F.}\ \bibnamefont {Maghrebi}},\
  }\bibfield  {title} {\bibinfo {title} {Nonequilibrium fixed points of coupled
  ising models},\ }\href@noop {} {\bibfield  {journal} {\bibinfo  {journal}
  {Physical Review X}\ }\textbf {\bibinfo {volume} {10}},\ \bibinfo {pages}
  {011039} (\bibinfo {year} {2020})}\BibitemShut {NoStop}%
\bibitem [{supplemental_material()}]{supplemental_material}%
  \BibitemOpen
  \bibinfo {note} {See supplemental materials for analytical derivations and
  mathematical details}\BibitemShut {NoStop}%
\bibitem [{\citenamefont {Szamel}(2014)}]{szamel2014self}%
  \BibitemOpen
  \bibfield  {author} {\bibinfo {author} {\bibfnamefont {G.}~\bibnamefont
  {Szamel}},\ }\bibfield  {title} {\bibinfo {title} {Self-propelled particle in
  an external potential: Existence of an effective temperature},\ }\href@noop
  {} {\bibfield  {journal} {\bibinfo  {journal} {Physical Review E}\ }\textbf
  {\bibinfo {volume} {90}},\ \bibinfo {pages} {012111} (\bibinfo {year}
  {2014})}\BibitemShut {NoStop}%
\bibitem [{\citenamefont {Sevilla}\ \emph {et~al.}(2019)\citenamefont
  {Sevilla}, \citenamefont {Rodr{\'\i}guez},\ and\ \citenamefont
  {Gomez-Solano}}]{sevilla2019generalized}%
  \BibitemOpen
  \bibfield  {author} {\bibinfo {author} {\bibfnamefont {F.~J.}\ \bibnamefont
  {Sevilla}}, \bibinfo {author} {\bibfnamefont {R.~F.}\ \bibnamefont
  {Rodr{\'\i}guez}},\ and\ \bibinfo {author} {\bibfnamefont {J.~R.}\
  \bibnamefont {Gomez-Solano}},\ }\bibfield  {title} {\bibinfo {title}
  {Generalized ornstein-uhlenbeck model for active motion},\ }\href@noop {}
  {\bibfield  {journal} {\bibinfo  {journal} {Physical Review E}\ }\textbf
  {\bibinfo {volume} {100}},\ \bibinfo {pages} {032123} (\bibinfo {year}
  {2019})}\BibitemShut {NoStop}%
\bibitem [{\citenamefont {Zamponi}\ \emph {et~al.}(2005)\citenamefont
  {Zamponi}, \citenamefont {Bonetto}, \citenamefont {Cugliandolo},\ and\
  \citenamefont {Kurchan}}]{zamponi2005fluctuation}%
  \BibitemOpen
  \bibfield  {author} {\bibinfo {author} {\bibfnamefont {F.}~\bibnamefont
  {Zamponi}}, \bibinfo {author} {\bibfnamefont {F.}~\bibnamefont {Bonetto}},
  \bibinfo {author} {\bibfnamefont {L.~F.}\ \bibnamefont {Cugliandolo}},\ and\
  \bibinfo {author} {\bibfnamefont {J.}~\bibnamefont {Kurchan}},\ }\bibfield
  {title} {\bibinfo {title} {A fluctuation theorem for non-equilibrium
  relaxational systems driven by external forces},\ }\href@noop {} {\bibfield
  {journal} {\bibinfo  {journal} {Journal of Statistical Mechanics: Theory and
  Experiment}\ }\textbf {\bibinfo {volume} {2005}},\ \bibinfo {pages} {P09013}
  (\bibinfo {year} {2005})}\BibitemShut {NoStop}%
\bibitem [{\citenamefont {Mandal}\ \emph {et~al.}(2017)\citenamefont {Mandal},
  \citenamefont {Klymko},\ and\ \citenamefont {DeWeese}}]{mandal2017entropy}%
  \BibitemOpen
  \bibfield  {author} {\bibinfo {author} {\bibfnamefont {D.}~\bibnamefont
  {Mandal}}, \bibinfo {author} {\bibfnamefont {K.}~\bibnamefont {Klymko}},\
  and\ \bibinfo {author} {\bibfnamefont {M.~R.}\ \bibnamefont {DeWeese}},\
  }\bibfield  {title} {\bibinfo {title} {Entropy production and fluctuation
  theorems for active matter},\ }\href@noop {} {\bibfield  {journal} {\bibinfo
  {journal} {Physical review letters}\ }\textbf {\bibinfo {volume} {119}},\
  \bibinfo {pages} {258001} (\bibinfo {year} {2017})}\BibitemShut {NoStop}%
\bibitem [{\citenamefont {Caprini}\ \emph {et~al.}(2019)\citenamefont
  {Caprini}, \citenamefont {Marconi}, \citenamefont {Puglisi},\ and\
  \citenamefont {Vulpiani}}]{caprini2019entropy}%
  \BibitemOpen
  \bibfield  {author} {\bibinfo {author} {\bibfnamefont {L.}~\bibnamefont
  {Caprini}}, \bibinfo {author} {\bibfnamefont {U.~M.~B.}\ \bibnamefont
  {Marconi}}, \bibinfo {author} {\bibfnamefont {A.}~\bibnamefont {Puglisi}},\
  and\ \bibinfo {author} {\bibfnamefont {A.}~\bibnamefont {Vulpiani}},\
  }\bibfield  {title} {\bibinfo {title} {The entropy production of
  ornstein--uhlenbeck active particles: a path integral method for
  correlations},\ }\href@noop {} {\bibfield  {journal} {\bibinfo  {journal}
  {Journal of Statistical Mechanics: Theory and Experiment}\ }\textbf {\bibinfo
  {volume} {2019}},\ \bibinfo {pages} {053203} (\bibinfo {year}
  {2019})}\BibitemShut {NoStop}%
\bibitem [{\citenamefont {Dabelow}\ \emph {et~al.}(2019)\citenamefont
  {Dabelow}, \citenamefont {Bo},\ and\ \citenamefont
  {Eichhorn}}]{dabelow2019irreversibility}%
  \BibitemOpen
  \bibfield  {author} {\bibinfo {author} {\bibfnamefont {L.}~\bibnamefont
  {Dabelow}}, \bibinfo {author} {\bibfnamefont {S.}~\bibnamefont {Bo}},\ and\
  \bibinfo {author} {\bibfnamefont {R.}~\bibnamefont {Eichhorn}},\ }\bibfield
  {title} {\bibinfo {title} {Irreversibility in active matter systems:
  Fluctuation theorem and mutual information},\ }\href@noop {} {\bibfield
  {journal} {\bibinfo  {journal} {Physical Review X}\ }\textbf {\bibinfo
  {volume} {9}},\ \bibinfo {pages} {021009} (\bibinfo {year}
  {2019})}\BibitemShut {NoStop}%
\end{thebibliography}


\clearpage
\newpage

\begin{widetext}

\section*{Supplemental Material}

This file is organized as follows: we first evaluate the correlation functions for a general $2$-dimensional system. This calculations will be useful for the derivations presented herein. Then, we discuss the $2$-dimensional case with reciprocal interactions, highlighting the role of a \textit{non-entropic} DOF. Later, we extend this to system composed of multiple \textit{non-entropic} DOFs. Finally, we deal with the case with non-reciprocal couplings, studying the effect of integrating out an \textit{entropic} DOF.

\section{Correlation functions for a general $2$D (linear) system}

\subsection{Equal-time correlation functions}

To estimate the correlation functions of a general $2$-dimensional system, start from the model in the main text:
\begin{equation}\label{eq:asym_2D_supp}
    \begin{split}
        & \dot{x}_t = A_{11}x_t+A_{12}y_t+\sqrt{2D}\xi^x_t \\
        & \dot{y}_t = A_{22}y_t+A_{21}x_t+\sqrt{2D}\xi^y_t.
    \end{split}
\end{equation}
Discretizing the dynamics, we obtain:
\begin{equation}\label{eq:asym_2D_supp_disc}
    \begin{split}
        & x_{t+dt} = x_{t} + (A_{11}x_t+A_{12}y_t)dt+\sqrt{2D}dW^x_t \\
        & y_{t+dt} = y_{t} + (A_{22}y_t+A_{21}x_t)dt+\sqrt{2D}dW^y_t.
    \end{split}
\end{equation}
where $dW^i_t$ is a Wiener process such that $\langle dW^i_t dW^j_t \rangle = \delta^{ij}dt$, with $i,j \in {x,y}$. Taking square and cross product of the equations in \eqref{eq:asym_2D_supp_disc}, then averaging and keeping only the terms of order $\mathcal{O}(dt)$, we have:
\begin{equation}\label{eq:asym_2D_supp_disc_square}
    \begin{split}
        & \langle x_{t+dt}^2\rangle = \langle x_{t}^2 \rangle  + 2(A_{11}\langle x^2_t\rangle+A_{12}\langle x_t y_t \rangle )dt+2Ddt \\
        & \langle y_{t+dt}^2\rangle = \langle y_{t}^2 \rangle  + 2(A_{22}\langle y^2_t\rangle+A_{21}\langle y_t x_t  \rangle )dt+2Ddt \\
         \langle x_{t+dt}y_{t+dt}&\rangle  = \langle x_{t}y_{t} \rangle  + (A_{11}\langle y_t x_t \rangle+A_{12}\langle y_t^2 \rangle + A_{22}\langle x_t y_t\rangle+A_{21}\langle x_t^2  \rangle )dt
    \end{split}
\end{equation}
where we used that $\langle a_t \, dW^i_t \rangle =0 $, $a=x,y$. By using that, in a steady state, $C_{\alpha\beta}(0) = C_{\beta\alpha}(0) = \langle \alpha_t \beta_t \rangle $ for every $t$, one sees that Eq. \eqref{eq:asym_2D_supp_disc_square} becomes the following linear system:
\begin{equation}\label{eq:asym_2D_supp_disc_square_2}
    \begin{split}
        &A_{11} C_{xx}(0)+A_{12} C_{xy}(0) = D\\[3pt]
        &A_{22} C_{yy}(0)+A_{21} C_{xy}(0) = D\\[3pt]
        A_{21}C_{xx}(0)&+A_{12}C_{xx}(0)+(A_{11}+A_{22})C_{xy}(0) =0
    \end{split}
\end{equation}
whose solution gives the equal time correlations:
\begin{equation}
\begin{split}\label{eq:asym_2D_eq_time_corr}
        & C_{xx}(0) = D\,\frac{A_{12} (A_{21} - A_{12}) - A_{22} (A_{11} + A_{22})}{(A_{11} + A_{22}) ( A_{11} A_{22}-A_{12} A_{21})}\\[5pt]
        & C_{yy}(0) = D\,\frac{A_{21} (A_{12} - A_{21}) - A_{11} (A_{11} + A_{22})) }{(A_{11} + A_{22}) (A_{11} A_{22}-A_{12} A_{21})}\\[5pt]
        & C_{xy}(0) = D\,\frac{A_{11} A_{12} + A_{21} A_{22}}{(A_{11} + A_{22}) (A_{11} A_{22} -A_{12} A_{21})}\\[4pt]
    \end{split}
\end{equation}

\subsection{Two-time correlation functions}

For time dependent correlations, by multiplying both equations in \eqref{eq:asym_2D_supp} by $x_t$ and $y_t$ respectively and averaging, one obtains:
\begin{equation}\label{eq:asym_2D_corr_time}
    \begin{split}
        & \partial_t C_{xx}(t) = A_{11}C_{xx}(t)+A_{12}C_{yx}(t) \\
        & \partial_t C_{xy}(t) = A_{11}C_{xy}(t)+A_{12}C_{yy}(t) \\
        & \partial_t C_{yx}(t) = A_{22}C_{yx}(t)+A_{21}C_{xx}(t)\\
        & \partial_t C_{yy}(t) = A_{22} C_{yy}(t)+A_{21} C_{xy}(t) \, ,
    \end{split}
\end{equation}
where we named $C_{\alpha\beta}(t)= \langle \alpha_t \beta _0 \rangle $ and used that $C_{a\xi}(t)=0$ for $a=x,y$. Performing a Laplace transform, we have:
\begin{equation}\label{eq:asym_2D_corr_laplace}
    \begin{split}
        & (s-A_{11}) C_{xx}(s) = C_{xx}(0) +A_{12}C_{yx}(s) \hspace{1.3cm} (s-A_{11}) C_{xy}(s) = C_{xy}(0)+ A_{12}C_{yy}(s) \\[3pt]
        & (s-A_{22}) C_{yx}(t) = C_{yx}(0)+A_{21} C_{xx}(s)\hspace{1.3cm} (s-A_{22}) C_{yy}(s) = C_{yy}(0) +A_{21} C_{xy}(s) \, ,
    \end{split}
\end{equation}
whose solution gives the correlation functions in the Laplace space:
\begin{equation}
\begin{split}\label{eq:asym_2D_corr_laplace_sol}
        & C_{xx}(s) = \frac{C_{xx}(0) (s-A_{22} )+A_{12} C_{xy}(0) }{(s-A_{11}  ) (s-A_{22})-A_{12} A_{21}}\hspace{1.3cm} 
        C_{xy}(s) = \frac{C_{xy}(0) (s-A_{22})+A_{12} C_{yy}(0) }{(s-A_{11}) (s-A_{22})-A_{12} A_{21}}\\[5pt]
        & C_{yx}(s) = \frac{C_{xy}(0) (s-A_{11})+A_{21} C_{xx}(0) }{(s-A_{11}) (s-A_{22})-A_{12} A_{21} }\hspace{1.3cm}
         C_{yy}(s) = \frac{C_{yy}(0) (s-A_{11})+A_{21} C_{xy}(0) }{(s-A_{11}) (s-A_{22} )-A_{12} A_{21}}\, .\\[4pt]
    \end{split}
\end{equation}
In the time domain, we have the following expressions:
\begin{eqnarray}
    C_{xx}(t) &=& e^{\frac{(A_{11}+A_{22})t}{2}} \left( C_{xx}(0) \cosh\left(t\Delta \right) + \frac{(A_{11} - A_{22})C_{xx}(0) + 2 A_{12} C_{xy}(0)}{2 \Delta} \sinh\left(t\Delta \right) \right) \nonumber \\
    C_{xy}(t) &=& e^{\frac{(A_{11}+A_{22})t}{2}} \left( C_{xy}(0) \cosh\left(t\Delta \right) + \frac{(A_{11} - A_{22})C_{xy}(0) + 2 A_{12} C_{yy}(0)}{2 \Delta} \sinh\left(t\Delta \right) \right) \\
    C_{yx}(t) &=& e^{\frac{(A_{11}+A_{22})t}{2}} \left( C_{yx}(0) \cosh\left(t\Delta \right) + \frac{(A_{22} - A_{11})C_{yx}(0) + 2 A_{12} C_{xx}(0)}{2 \Delta} \sinh\left(t\Delta \right) \right) \nonumber \\
    C_{yy}(t) &=& e^{\frac{(A_{11}+A_{22})t}{2}} \left( C_{yy}(0) \cosh\left(t\Delta \right) + \frac{(A_{22} - A_{11})C_{yy}(0) + 2 A_{12} C_{yx}(0)}{2 \Delta} \sinh\left(t\Delta \right) \right) \nonumber
\end{eqnarray}
where $\Delta =  \sqrt{ A_{12} A_{21} + ((A_{11} - A_{22})/2)^2}$.

\section{$2$D Case with reciprocal couplings}

\subsection{Entropy production rate at stationarity}

Here, we show the calculations involved in the treatment of the system in Eq.~(1) of the main text, considering $A_{12}=A_{21}$ and $f_y = 0$ to simplify the notation. The total entropy production rate at stationarity is:
\begin{gather}
    \sigma_{\rm tot} = \langle F^x_t \circ \dot{x}_t \rangle + \langle F^y_t \circ \dot{y}_t \rangle = \langle f^x_t \circ \dot{x}_t \rangle + \langle A_{12} y_t \circ \dot{x}_t \rangle + \langle A_{11} x_t \circ \dot{x}_t \rangle + \langle A_{21} x_t \circ \dot{y}_t \rangle + \langle A_{22}y_t\circ\dot{y}_t\rangle = \nonumber \\[4pt]
    = \langle f^x_t \circ \dot{x}_t \rangle + \Bigl\langle \frac{\partial V}{\partial x}\circ\dot{x}_t\Bigl\rangle + \Bigl\langle \frac{\partial V}{\partial y}\circ\dot{y}_t\Bigl\rangle = \langle f^x_t \circ \dot{x}_t \rangle + \frac{\partial}{\partial t} \langle V \rangle = \langle f^x_t \circ \dot{x}_t \rangle,
\end{gather}
where $V(x_t,y_t)= A_{11}x_t^2/2+ A_{12} x_t y_t+A_{22}y_t^2/2$ is the potential that generates the reciprocal interactions.

The entropy production associated with the $x$ DOF is obtained from the equations of motion upon integration of $y_t$. One gets:
\begin{eqnarray}
        \sigma_{x} &=& ~ \langle F^x_t \circ \dot{x}_t \rangle = \langle f^x_t \circ \dot{x}_t \rangle + \langle A_{11} x_t\circ\dot{x}_t \rangle + A_{12}A_{21}\int_0^t\mathrm{d}s\, e^{A_{22}(t-s)} \langle x_s\circ\dot{x}_t \rangle+A_{12}\sqrt{2D}\int_0^t\mathrm{d}s\,e^{A_{22}(t-s)} 
        \langle \xi_s^y\circ\dot{x}_t \rangle = \nonumber \\
        \label{sigmaX}
        &=& ~\langle f^x_t \circ \dot{x}_t \rangle + A_{12}A_{21}\int_0^t\mathrm{d}s\, e^{A_{22}(t-s)} \partial_t C_{xx}(t-s)-A_{12}\int_0^t\mathrm{d}s\,e^{A_{22}(t-s)}\partial^2_tC_{xy}(t-s) - \\
        &~& -A_{12}A_{22}\int_0^t\mathrm{d}s\,e^{A_{22}(t-s)}\partial_tC_{xy}(t-s)-A_{12}A_{21}\int_0^t\mathrm{d}s\,e^{A_{22}(t-s)}\partial_tC_{xx}(t-s) = \langle f^x_t \circ \dot{x}_t \rangle \;, \nonumber
\end{eqnarray}
where this equality has been obtained by expressing $\xi^y_s$ in terms of the dynamics of $y_s$, and the final result just replacing the form of the correlation functions and performing the integrals, considering that $A_{12} = A_{21}$.

\subsection{Effective GLE and its entropy production rate}

Considering $x_t$ as the observed DOF, we can integrate out $y_t$ exactly. By solving the dynamics of $y_t$, we obtain the following formal solution:
\begin{equation}
    y_t = y_{0} e^{A_{22}(t-t_0)} + \int_{t_0}^t \mathrm{d}s\, e^{A_{22}(t-s)}\left[A_{12} x_s+\sqrt{2D}~\xi_s^y \right].
\end{equation}
Since we are interested in the steady-state, we take $t_0 \to -\infty$, so that we have the following resulting equation for $x_t$:
\begin{equation}
    \dot{x}_t = f^x_t + A_{11}x_t + A_{12}\Bigl\{ \int_{-\infty}^t \mathrm{d}s ~e^{A_{22}(t-s)}\left[A_{12} x_s+\sqrt{2D}~\xi_s^y \right]\Bigl\}+\sqrt{2D}~\xi_t^x.
\end{equation}
We immediately notice that there is an additional noise term which is exponentially correlated. This can be incorporated into the Gaussian white noise. Indeed, the reconstruction of the corresponding GLE is done by identifying $\int_{-\infty}^t \mathrm{d}s\, e^{A_{22}(t-s)}\sqrt{2D}\xi_s^y+\sqrt{2D}\xi_t^x$ as a single Gaussian noise term, $\eta^x_t$, with correlation function (following the supplementary material of \cite{OurGLE}):
\begin{equation}
    \langle\eta_t\eta_{s}\rangle = D\left(2\delta(t-s)-\frac{A_{12}^2}{A_{22}}e^{A_{22}|t-s|}\right) = D \Gamma(t-t').
\end{equation}
The corresponding GLE kernel $\Gamma(t-t')$ is constructed by requiring the fluctuation-dissipation theorem to be satisfied (as for the original system). This can either be done by introducing by hand a term $\int_{-\infty}^t\mathrm{d}s\,\Gamma(t-s)\dot{x}_s$ or by extracting the correct dissipation kernel from the introduced memory term $A_{12}^2\int_{-\infty}^t \mathrm{d}s\, e^{A_{22}(t-s)} x_s$ through an integration by part. In both cases, the corresponding GLE for the observed DOF, $x_t$, reads:
\begin{equation}
    \int_0^t \mathrm{d}s\, \Gamma(t-s)\dot{x}_s = f_x+A_{11}x_t - \mathcal{A}^{12} x_t + \eta^x_t \;,
    \label{eq:SGLEx}
\end{equation}
where $\mathcal{A}^{12} = A_{12}^2/A_{22}$. Since the explicit integration of $y_t$ introduces only potential terms in the effective equation for $x_t$, by applying the customary Sekimoto's formula using $f^x_t + A_{11} x_t - \mathcal{A}^{12} x_t$ as a force, we have:
\begin{equation}
    \sigma_{\rm GLE}^{(x)} = \langle \left( f^x_t + A_{11} x_t - \mathcal{A}^{12} x_t \right) \circ \dot{x}_t \rangle = \langle f^x_t \circ \dot{x}_t \rangle + \frac{1}{2} \left( A_{11} - \mathcal{A}^{12} \right) \partial_t C_{xx}(0) = \langle f^x_t \circ \dot{x}_t \rangle
\end{equation}
The entropy production rate of $x_t$ at stationarity, as calculated from the GLE \eqref{eq:SGLEx}, corresponds to $\sigma_{tot} = \sigma_x$.

\section{System with multiple \textit{non-entropic} DOFs}

We now focus on the case in which some observed DOFs are coupled to multiple \textit{non-entropic} DOFs. Consider that we already integrated out all but one of them to obtain an effective GLE for the evolution of $x^i_t$. The remaining coupled equations, also reported in the main text, are:
\begin{equation}\label{eq:complete_LE}
\begin{split}
    \int_{-\infty}^{t} \mathrm{d}s \,\Gamma^{ij}(t-s)\dot{x}^j_s = f^{i}_{t} + A_{xy}^{i}y_t + \eta^{i}_t \\
   \dot{y}_t = A_{yy} y_t + A_{xy}^{i} x^i_t + \sqrt{2D}\, \xi^y_t \, ,
\end{split}
\end{equation}
where Einstein's notation is used, $\langle \eta^i_{t'} \eta^j_{t'} \rangle = \Gamma^{ij}(|t'-t''|)$, $\langle \xi^y_{t'} \xi^y_{t''} \rangle = \delta(t'-t'') $ and where $f^i_t$ is a generic non-conservative force accountable for a non-zero entropy production rate at stationarity. We can understand this by defining $\pmb{r}=(\pmb{x},y)$ and directly applying the Sekimoto's formula:
\begin{equation}\label{eq:EP_complete}
\begin{split}
    \sigma = \langle \pmb{F}_t \circ \dot{\pmb{r}}_t \rangle =& \langle f^i_t \circ \dot{x}^i_t \rangle + A^i_{xy}\langle y_t \circ \dot{x}^i_t \rangle +A_{yy}\langle y_t \circ \dot{y}^i_t \rangle + A^i_{xy}\langle x^i_t \circ \dot{y}_t \rangle \\
    = & \langle f^i_t \circ \dot{x}^i_t \rangle + \partial_t( A^i_{xy} \langle x^i_t y_t \rangle + A_{yy} \langle y_t y_t \rangle/2  )\\
    = & \langle f^i_t \circ \dot{x}^i_t \rangle \, .
\end{split}
\end{equation}
This results coincides with $\sigma_x$, in analogy to the previous $2$-dimensional case. This suggests that $y_t$ can be seen as a thermal bath absorbing and releasing the same amount of heat into the \textit{entropic} observed variables. Indeed, we will see that Sekimoto's formula for the entropy production rate remains valid and gives the same result if we integrate out the variable $y_t$ and obtain an effective GLE for the $x^i_t$ degrees of freedom. To do so, we formally solve the second equation in \eqref{eq:complete_LE}, as before:
\begin{equation}
    y_t = y_0 e^{A_{yy}(t-t_0)
    } + \int_{t_0}^t\mathrm{d}s\, e^{A_{yy}(t-s)}\left(A_{xy}^{i} x^i_s+\sqrt{2D}\,\xi^y_s\right) \, ,
\end{equation}
where $t_0$ is the initial time for the beginning of the dynamics. Since we are interested in the steady state, we set $t_0 \to - \infty$:
\begin{equation}
    y_t = \int_{-\infty}^t\mathrm{d}s\, e^{A_{yy}(t-s)}\left(A_{xy}^{i} x^i_s +\sqrt{2D}\,\xi^y_s\right) \, .
\end{equation}
By plugging this result into the first of equations \eqref{eq:complete_LE}, we get the following effective GLE:\\
\begin{equation}\label{eq:substitution}
\begin{split}
    \int_{-\infty}^{t} \mathrm{d}s \,\Gamma^{ij}(t-s)\dot{x}^i_s &= f^{i}_{t} + A_{xy}^{i}A_{xy}^{j} \int_{-\infty}^t\mathrm{d}s\, e^{A_{yy}(t-s)} x^j_s + \widetilde{\eta}^{\,i}_t \\
    \widetilde{\eta}^{\,i}_t &= A_{xy}^{i}+\int_{-\infty}^t\mathrm{d}s\, e^{A_{yy}(t-s)}\xi^y_s \, .
\end{split}
\end{equation}
As before, we can identify a new colored Gaussian noise term whose correlation can be again calculated as in the supplementary material in \cite{OurGLE} and are equal to:
\begin{equation}
    \langle \widetilde{\eta}^{\,i}_{t'} \widetilde{\eta}^{\,j}_{t''} \rangle = \left( \Gamma^{ij}(|t'-t''|) -\mathcal{A}^{ij}e^{A_{yy}|t'-t''|}\right) = \widetilde{\Gamma}^{\,ij} (|t'-t''|) \, ,
\end{equation}
where $\mathcal{A}^{ij}=A_{xy}^{i}A_{xy}^{j}/A_{yy}$. Furthermore, by integrating by parts the first equation in \eqref{eq:substitution} one gets:\\
\begin{equation}\label{eq:GLE_final}
    \int_{-\infty}^{t} \mathrm{d}s \,\widetilde{\Gamma}^{\,ij}(t-s)\dot{x}^i_s = f^{i}_{t} -\mathcal{A}^{ij}x^{j}_t+\widetilde{\eta}^{\,i}_t \, .
\end{equation}
We recovered a GLE following the fluctuation dissipation theorem. As such, the system can be displaced from equilibrium only due to the non conservative force. Indeed, by applying the Sekimoto's formula to \eqref{eq:GLE_final}, i.e., the force is $f^i_t - \mathcal{A}^{ij}x^j_t$, we obtain: 
\begin{equation}\label{eq:EP_GLE}
    \begin{split}
        \sigma_{\rm GLE}^{(x)} =&  \langle (f^i_t-\mathcal{A}^{ij}x^{j}_t) \circ \dot{x}^i_t \rangle \\[5pt]
        = & \langle f^i_t \circ \dot{x}^i_t \rangle - \mathcal{A}^{ij} \langle x^{j}_t \circ \dot{x}^i_t \rangle \\[5pt]
        = & \langle f^i_t \circ \dot{x}^i_t \rangle  - \frac{\mathcal{A}^{ij}}{2dt} \langle (x^{j}_{t+dt}+x^{j}_{t})(x^{i}_{t+dt}-x^{i}_{t}) \rangle \\[5pt]
        = & \langle f^i_t \circ \dot{x}^i_t \rangle  - \frac{\mathcal{A}^{ij}}{2dt} \langle C_{xx}^{ij}(dt)-C_{xx}^{ji}(dt) \rangle \\[5pt]
        = & \langle f^i_t \circ \dot{x}^i_t \rangle - \mathcal{A}^{ij}\langle \dot{C}_{xx}^{ij}(0)-\dot{C}_{xx}^{ji}(0) \rangle/2 \\[5pt]
        =&\langle f^i_t \circ \dot{x}^i_t \rangle = \sigma_{x} = \sigma_{\rm tot} \, ,
    \end{split}
\end{equation}
where $C_{xx}^{\,ij}(t) = \langle x^i_t x^j_0\rangle$ and $\dot{C}_{xx}^{ij}(0)=\partial_t C_{xx}^{\,ij}(t)|_{t=0}$. This proves that the GLE provides a thermodynamically consistent description for observed variables coupled to \textit{non-entropic} DOFs.

\section{$2$D (General) Case with non-reciprocal couplings}

\subsection{Entropy production rate at stationarity}

Here we consider the behaviour of the dynamics of a generic two-dimensional system with non-reciprocal couplings under integration of one degree of freedom. The starting point is the system of coupled equations \eqref{eq:asym_2D_supp_disc}:
\begin{equation}
    \begin{split}
        & \dot{x}_t = A_{11}x_t+A_{12}y_t+\sqrt{2D}\xi^x_t \\
        & \dot{y}_t = A_{22}y_t+A_{21}x_t+\sqrt{2D}\xi^y_t.
    \end{split}
\end{equation}
The entropy production associated with the whole system is calculated again using Sekimoto's formula and gives:
\begin{equation}
    \sigma_{tot} = \langle F_x\circ\dot{x}_t \rangle + \langle F_y\circ\dot{y}_t \rangle = A^2_{12} C_{yy} + A^2_{21} C_{xx} + \left( A_{11}A_{12} + A_{22} A_{21} \right) C_{xy} = - \frac{(A_{12} - A_{12})^2}{A_{11} + A_{22}},
    \label{eqS:totalep}
\end{equation}
where $C_{xx}(0) = \langle x_t x_t \rangle$, $C_{yy} = \langle y_t y_t \rangle$, $C_{xy}(0) = \langle x_t y_t \rangle$ in steady state. This result can be immediately obtained by plugging into the resulting equation the expressions of the variances derived above.

\subsection{Entropy production rate for the observed DOF only}

As in the case of reciprocal couplings, formal integration of $y$ gives:
\begin{equation}
    y_t = A_{21}\int_{-\infty}^t \mathrm{d}s\, e^{A_{22}(t-s)}x_s + \sqrt{2D}\int_{-\infty}^t \mathrm{d}s\, e^{A_{22}(t-s)}\xi_s^y.
\end{equation}
As a consequence, the final equation for $x_t$ is then:
\begin{equation}
    \dot{x}_{t} = A_{11}x_t + A_{12}A_{21}\int_{-\infty}^t \mathrm{d}s\, e^{A_{22}(t-s)}x_s+A_{12}\sqrt{2D}\int_{-\infty}^t \mathrm{d}s\,  e^{A_{22}(t-s)}\xi_s^y + \sqrt{2D}\xi_t^x.
    \label{eqS:xinty}
\end{equation}
By identifying as a force all the terms but the intrinsic noise of $x_t$, the entropy production rate associated with $x_t$ can be immediately calculated using the customary Sekimoto's formula. The steps are the same as in Eq.~\eqref{sigmaX} and we get:
\vspace{0.25cm}
\begin{equation}
    \sigma_x = \langle F_x\circ\dot{x}_t \rangle = \left\langle \left( \dot{x}_t - \sqrt{2D}~\xi^x_t \right) \circ \dot{x}_t \right\rangle = \frac{A_{12} \left( A_{21} - A_{12} \right)}{A_{11} + A_{22}}
\end{equation}
which is different from zero as $A_{12} \neq A_{21}$. Notice that, as shown above, this goes to $0$ when $A_{12}=A_{21}$ (reciprocal couplings) since there are no non-conservative forces in the model we are considering. $\sigma_x$ corresponds to the contribution to the total entropy production calculated in \eqref{eqS:totalep} due to $x_t$ only.

\subsection{Effective GLE and its entropy production rate}

In order to reconstruct an effective bath resulting from the influence of $y_t$ to $x_t$, we rearrange the terms of Eq.~\eqref{eqS:xinty} to obtain an effective GLE satisfying the fluctuation-dissipation theorem. This procedure is analogous to the one presented above. It first leads to the following equation:
\begin{equation}
    \dot{x}_t = A_{11}x_t + A_{12}A_{21}\int_0^t \mathrm{d}s\, e^{A_{22}(t-s)}x_s + \eta^x_t,
\end{equation}
where $\eta_t$ is a Gaussian noise with correlations (estimated again following \cite{OurGLE}):
\begin{equation}
    \langle \eta^x_t\eta^x_{t'} \rangle = D\left(2\,\delta(t-t') - \mathcal{A}^{12}e^{A_{22}|t-t'|}\right) = D \Gamma(t-t').
    \label{eqS:noisemem}
\end{equation}
where we introduce $\mathcal{A}^{12} = A_{12} A_{21}/A_{22}$. The corresponding dissipative kernel is obtained by integrating by parts the term with memory in the $x_t$ dynamics and merging with $\dot{x}_t$ only the contribution which satisfies the fluctuation-dissipation theorem with the noise \eqref{eqS:noisemem}. This procedure then gives:
\begin{equation}
    \int_0^t \mathrm{d}s\,\Gamma(t-s)\dot{x}_s = A_{11}x_t + A_{12}(A_{21}-A_{12})\int_0^t \mathrm{d}s\, e^{A_{22}(t-s)}x_s - \mathcal{A}^{12} x_t + \eta_t^x,
\end{equation}
with $\Gamma(t-t') = 2\delta(t-t') - \mathcal{A}^{12} e^{A_{22}|t-t'|}$. We can write this dynamics in a reduced form, as in the main text, by introducing a rescaled coupling $\tilde{A}_{11}$ and the kernel $K_f$ defined as follows:
\begin{equation}
    \tilde{A}_{11} = A_{11} - \mathcal{A}^{12} \qquad \qquad \qquad K_f(\tau) = A_{12} (A_{21} - A_{12}) e^{A_{22} \tau}
\end{equation}
The entropy production can be again calculated by noting that the system is driven by a force with memory and applying the Sekimoto's formula (that provided thermodynamically consistent results in the case of reciprocal couplings):
\begin{equation}
    \sigma_{GLE}^{(x)} = \left\langle A_{12}(A_{21}-A_{12})\int_0^t \mathrm{d}s\, e^{A_{22}(t-s)}x_s \circ \dot{x}_t \right\rangle = \sigma_x \left( 1 + \frac{\alpha^{-1} \mathcal{A}^{12}}{2 \left(A_{11} + A_{22}\right) - \mathcal{A}^{12}} \right),
\end{equation}
by using the explicit form of the correlation functions derived above and performing the integrals, where $\alpha$ is defined by $A_{21} = \alpha A_{12}$ and quantifies the non-reciprocity of interactions.

\end{widetext}

\end{document}